\documentclass[reprint, nofootinbib, aps, preprintnumbers]{revtex4-2}
\usepackage{color}
\usepackage{slashed}
\usepackage[utf8]{inputenc}
\usepackage[toc,page]{appendix}
\usepackage{graphicx}
\usepackage{placeins}
\usepackage{subfigure}
\usepackage{booktabs}
\usepackage{comment}
\usepackage{dcolumn}
\usepackage{physics}
\usepackage{float}
\usepackage{bbold}
\usepackage{amsmath,amsfonts,bbm,mathrsfs,amssymb}
\usepackage{slashed,cancel}
\usepackage[usenames,dvipsnames]{xcolor}
\usepackage{catchfile}
\usepackage{indentfirst}
\usepackage{enumerate}
\usepackage{multirow}
\usepackage{stmaryrd} 
\usepackage{pifont}% http://ctan.org/pkg/pifont
\usepackage{soul}
\usepackage{ragged2e}
\usepackage{bbm}
\usepackage{bm}
\usepackage{comment}
\usepackage{framed}
 \definecolor{shadecolor}{named}{SpringGreen}

\DeclareGraphicsExtensions{.pdf,.png,.jpg}

\usepackage[colorlinks=true, linkcolor=blue, citecolor=blue,urlcolor=blue,bookmarks=true,bookmarks=true,bookmarksopen=true,bookmarksnumbered=true,bookmarksopenlevel=3]{hyperref}

\def \thl {{\theta_l}}
\def \thK {{\theta_{K^*}}}
\definecolor{airforceblue}{rgb}{0.36, 0.54, 0.66}
\definecolor{steelblue}{rgb}{0.27, 0.51, 0.71}
\definecolor{amber}{rgb}{1.0, 0.49, 0.0}
\definecolor{darkgreen}{rgb}{0.0, 0.5, 0.0}
\definecolor{amber}{rgb}{1.0, 0.49, 0.0}

\DeclareMathAlphabet{\mathpzc}{OT1}{pzc}{m}{it}

\graphicspath{{draft/Figures/}}

\def\simg{{\ \lower-1.2pt\vbox{\hbox{\rlap{$>$}\lower6pt\vbox{\hbox{$\sim$}}}}\ }}
\def\siml{{\ \lower-1.2pt\vbox{\hbox{\rlap{$<$}\lower6pt\vbox{\hbox{$\sim$}}}}\ }}
\newcommand*{\Scale}[2][4]{\scalebox{#1}{$#2$}}

\renewcommand{\theequation}{\arabic{equation}}

\newcommand*{\av}[1]{\langle #1 \rangle}

\newcommand\subsetsim{\mathrel{%
  \ooalign{\raise0.2ex\hbox{$\subset$}\cr\hidewidth\raise-0.8ex\hbox{\scalebox{0.9}{$\sim$}}\hidewidth\cr}}}

\graphicspath{{Figures/}}

\begin{document}

\preprint{IFT-UAM/CSIC-22-139}
\title{
Neutral $B$-anomalies from an \textit{on-shell} scalar exchange}

\author{J. Bonilla}\email{jesus.bonilla@uam.es}
\author{A. de Giorgi}\email{arturo.degiorgi@uam.es} \author{M. Ramos}\email{maria.pestanadaluz@uam.es}
\affiliation{
 Departamento de F\'isica Te\'orica and Instituto de F\'isica Te\'orica UAM/CSIC,\\
Universidad Aut\'onoma de Madrid, Cantoblanco, 28049, Madrid, Spain
}

%  Abstract
 \begin{abstract}
The neutral $B$-anomalies are analysed in terms of the tree-level exchange of a (pseudo)scalar gauge singlet $a$. Solutions to both $R_{K^{(*)}}$ central bin anomalies are found within $1\sigma$ for ${m_{a}^2 \in [{1.1},\,{6}]\,\text{GeV}^2}$, while the low $q^2$-bin anomaly can also be accounted for with masses close to the bin threshold. The impact of these solutions on other ${b\to s e^+ e^-}$ observables is discussed in detail. Due to the \textit{on-shell} enhancement, sizable effects are expected in null tests of the SM, such as the flat term, $F_H$, of the $B\to K e^+ e^-$ angular distribution. At the same time, the observable sensitive to the $K^\ast$ polarisation, $F_L\, (B\to K^* e^+ e^-)$, and the lepton forward-backward asymmetry, ${A_{FB}\, (B\to K^* e^+ e^-)}$, can be suppressed with respect to their SM values. Corrections from the new physics to $\mathcal{B}(B_s \to e^+ e^-)$ are, on the other hand, negligible. Along with the previous observables, improved measurements of the cross section ${\sigma (e^+ e^- \to a (e^+ e^-) \gamma)}$ could potentially probe the relevant parameter space of the model. A comparison between our results and those stemming from an axion-like particle exchange is also discussed, showing that the exchange of a general scalar singlet offers a noticeably wider parameter space.
\end{abstract}

%%%%%%%%%
\maketitle
%%%%%%%%%%%%%%%%%%%%

\section{Introduction}
In the last years, a pattern of significant deviations with respect to the Standard Model (SM) predictions has emerged on measurements of $b\to s \ell^+ \ell^-$ observables. These include the lepton-flavour universality (LFU) violating ratio ${R_{K^*}} \equiv \frac{\mathcal{B} (B\to K^* \mu^+ \mu^-)}{\mathcal{B} (B\to K^* e^+ e^-)}$ measured by the LHCb collaboration in two bins of the di-lepton invariant mass squared, $q^2 \in [0.045,\,1.1]$ and $[1.1,\,6.0]\,\text{GeV}^2$, as well as ${{R_{K}} \equiv \frac{\mathcal{B} (B\to K \mu^+ \mu^-)}{\mathcal{B} (B\to K e^+ e^-)}}$ measured only in the latter {bin}. Given the recent data published by the collaboration~\cite{LHCb:2021trn}, $R_K$ shows the largest tension with the SM, at the level of $3.1\sigma$; while the deviations in the low and central bins of $R_{K^{*}}$ are at the $2.3\sigma$ and $2.5\sigma$ level, respectively~\cite{LHCb:2017avl}. Results from other $B$ factories suffer from larger uncertainties~\cite{Belle:2009zue, BaBar:2012mrf} and are in turn consistent with the SM predictions.

Being computed as double ratios, the hadronic and systematic uncertainties cancel in those observables to large extent~\cite{Hiller:2003js, Bobeth:2007dw, Bouchard:2013mia}, making them an ideal place to search for BSM effects in either of the lepton channels. Notwithstanding, the fact that other deviations have been observed in the muon sector has propelled analyses within the SM effective field theory (SMEFT) framework including new physics (NP) coupled only to  muons~\cite{Alguero:2019ptt,Alok:2019ufo,Ciuchini:2019usw,DAmico:2017mtc,Aebischer:2019mlg, Guedes:2022cfy}, that must interfere destructively with the SM in order to make $R_{K^{(*)}} < 1$ (and hence be of the vector/axial form). Some of these deviations were found in angular observables of the $B \to K^* \mu^+ \mu^- $ distribution~\cite{LHCb:2015svh,Belle:2016xuo} and in the branching ratio $\mathcal{B}(B\to \phi \mu^+ \mu^-)$~\cite{LHCb:2021zwz}. Depending on the assumptions on the size of hadronic corrections, the statistical significance of the experimental measurements can, however, vary from $\sim 4 \sigma$ to less than $2\sigma$~\cite{Ciuchini:2015qxb,Hurth:2016fbr,Chobanova:2017ghn,Hurth:2020rzx}. Following a conservative approach, we are therefore led to consider only the cleanest $b\to s \ell^+ \ell^-$ observables, in which case NP affecting solely the muon sector is as plausible as NP in the electron one (for a combined analysis of the two channels within the SMEFT approach, see \textit{e.g.} Ref.~\cite{Kumar:2019qbv}).

On top of that, global SMEFT fits to relevant data on $b\to s$ transitions, while being able to account for the central $R_{K^{(*)}}$ bin anomalies, cannot solve satisfactorily the tension in the low bin. 
 Indeed, at low $q^2$ the semileptonic rates are dominated by the photon pole contribution, diluting significantly the effects from contact NP interactions~\cite{Altmannshofer:2017yso,Capdevila:2017bsm,DAmico:2017mtc,Geng:2017svp,Ciuchini:2017mik}. This has motivated interesting studies where light \textit{d.o.f.} were included in the phenomenological analysis~\cite{Datta:2017pfz,Sala:2017ihs,Ghosh:2017ber,Alok:2017sui,Bishara:2017pje,Datta:2017ezo,Altmannshofer:2017bsz}. If such light physics contributes to the $B$-anomalies via an \textit{on-shell} exchange, it can only suppress the LFU ratios via the electron mode. Such possibility was analysed recently in Ref.~\cite{Bonilla:2022qgm} within the axion-like particle (ALP) EFT framework~\cite{Brivio:2017ije,Alonso-Alvarez:2018irt,Chala:2020wvs,Bonilla:2021ufe}. It was found however that, due to the derivative nature of the ALP interactions, solutions to the anomalies require very large couplings to leptons, limiting the region of the phase space compatible with the EFT validity.

In this work, we argue that the tree-level exchange of a generic scalar singlet, comprising both shift-symmetric and shift-breaking interactions, would ameliorate these problems. We furthermore study the impact of the anomaly solutions on several other flavour observables, that can be enhanced \textit{on resonance}. These include the differential semileptonic and leptonic $B$-decay rates, the fraction of the longitudinal polarization of the $K^*$-meson ($F_L$) emitted in $B \to K^\ast e^+ e^-$ decays and the forward-backward asymmetry ($A_{FB}$) of the di-lepton system, among others. The former has been measured in the $q^2 \in [0.0002,1.12]\,\text{GeV}^2$ bin, along with other angular observables such as ${P_i^{e\prime}}$. On the contrary, no data is currently available on the angular $B\to K e^+ e^-$ distribution. By studying the latter, we can therefore quantify the potential of new experimental analyses in probing the scalar setup.

We also estimate the NP contribution to other LFU ratios, such as $R_\phi \equiv \frac{\mathcal{B}(B_s\to \phi\mu^+\mu^-)}{\mathcal{B}(B_s\to \phi e^+e^-)}$, that is planned to be determined in the next LHCb run in both $q^2\in[0.1,1.1]$ and $[1.1,6.0]\,\text{GeV}^2$ bins. The prospects to constrain the model parameter space at  electron-positron colliders are also discussed.

Moreover, we explore the impact on data of the exchange of a singlet with mass $\mathcal{O}(1)$ GeV that can accommodate all low and central bin anomalies of $R_{K^{(*)}}$. While this \textit{golden} solution was identified in Ref.~\cite{Bonilla:2022qgm}, we assess its statistical significance by performing a $\chi^2$-fit to the experimental data and quantify how much better the NP is at explaining those than the SM. We will also compare the preferred regions of the singlet parameter space with that of an ALP.

This article is organised as follows. In section~\ref{sec:L}, we present the theoretical framework and define the region of the parameter space of phenomenological interest. In section~\ref{sec:pheno}, we give predictions for the $b\to s \ell^+ \ell^-$ observables mostly impacted by the new resonance exchange. The fit to the three anomaly solutions is presented in section~\ref{sec:golden}. Finally, section~\ref{sec:conc} is dedicated to our conclusions. The relevant experimental constraints and the SM inputs used throughout this work are presented in App.~\ref{sec:obs} and~\ref{app:SM}, as well as some useful relations in App.~\ref{app:on-shell}. In App.~\ref{sec:offshell} we make a brief comparison between the off-shell and the on-shell results obtained in this work.

\section{(pseudo)scalar new physics}~\label{sec:L}
At the renormalizable level, interactions between a gauge singlet and the Higgs boson can induce couplings between the former and the SM fermions after electroweak symmetry breaking (EWSB). We neglect such couplings in our analysis since \textit{(i)} they are not only suppressed by the lepton masses, but also by the small singlet-Higgs mixing angle~\cite{Clarke:2013aya}; and \textit{(ii)} such couplings generate flavour-violating (FV) effects -- including those required to produce the $bs$-singlet vertex -- but only at one-loop level via a $W$ exchange. Since in the SM the semileptonic rates of interest are also generated at this order in perturbation theory, and we aim to account for a departure with respect to the SM predictions, we assume that FV couplings are present at tree level in the quark singlet Lagrangian below the EW scale. It is beyond the scope of our work to construct a complete UV model from where these effects could arise. However, we remark that  flavour changing neutral currents in the quark sector can arise naturally in the composite Higgs framework~\cite{Gripaios:2009pe,Chala:2016ykx,Balkin:2017aep,DaRold:2019ccj,Blance:2019ixw}, mediated by exotic gauge singlets predicted in several non-minimal symmetry breaking patterns~\cite{Bellazzini:2014yua}.

For simplicity, we do not consider couplings of the singlet to gauge bosons or to itself that do not have a leading effect in the following analysis; we further assume stability in the scalar potential along the singlet direction. Along the work, we will compare the phenomenology 
of a CP-even ($a_+$) vs. a CP-odd ($a_-$) light scalar field. The most general and minimal Lagrangian that encodes their interactions with the SM fermions reads:
\begin{equation}
    \mathscr{L}^{\rm int}_{a_{\pm}} = \frac{ a_{\pm}}{\Lambda} (\mathbf{C}_{\psi}^{\pm})_{\alpha \beta} \left( \overline{\Psi_L^\alpha} {\Phi} \Psi_R^\beta \pm \overline{\Psi_R^\beta} {\Phi^\dagger} \Psi_L^\alpha \right)\,,
\end{equation}
where $\Phi$ denotes the Higgs doublet (or ${\widetilde{\Phi}\equiv \text{i} \sigma_2 \Phi^\ast}$, in case of interactions with the up-type quarks), $\Psi_{L} \in \{Q_L,\, L_L\}$ and $\Psi_R\in\{ u_R,\,d_R,\,e_R\}$ are the $SU(2)_L$ doublet and singlet SM matter fields, with $\psi$ denoting their flavour ($\psi \in \{ u \,, d \,, e \,, \nu\}$), and $\Lambda$ is the UV cutoff scale. We assume that CP is a conserved symmetry of the NP sector, hence the flavour matrices $\mathbf{C}^+$ and $\text{i}\mathbf{C}^-$ are real. We work under the assumption that the couplings $\mathbf{C}^\pm$ stay below the maximal value of $(4\pi)^2$ allowed by perturbativity in the UV theory~\cite{Gavela:2016bzc}. Furthermore, for the EFT to be predictive, the condition $\mathbf{C}^\pm\,\text{max}( \sqrt{q^2}/\Lambda,\,m_{a_\pm}/\Lambda) < 1$ must be satisfied.

As the phenomenology of our interest occurs well below EWSB, we absorb the dimension-five effects into the following renormalizable interactions:
\begin{align}
    \mathscr{L}^{\rm int}_{a_+} &=  {a_+} (C_{\psi}^+)_{\alpha \beta} \left( \overline{\psi_L^\alpha} \psi_R^\beta + \overline{\psi_R^\beta} \psi_L^\alpha \right)+\dots\,;\\
    \mathscr{L}^{\rm int}_{{a_-}} &= \text{i} {a_-} (C_{\psi}^{-})_{\alpha \beta} \left( \overline{\psi_L^\alpha}  \psi_R^\beta - \overline{\psi_R^\beta} \psi_L^\alpha \right)+\dots\,,
\end{align}
with $C^\pm \equiv (-\text{i})\mathbf{C}^\pm v/ (\sqrt{2}\Lambda)$. The dots include interactions with the Higgs boson that are outside the scope of the present work.

In the $m_{a_\pm} \gg m_B$ regime, where $m_{a_\pm}$ and $m_B$ denote, respectively, the singlet and $B$-meson masses, the singlet can be integrated out sourcing the flavour changing operators
\begin{align}
    \mathcal{O}_S^{(')} = \left(\overline{s} P_{R(L)}  b\right) \left(\overline{\ell}  \ell\right);~ \mathcal{O}_P^{(')}  = \left(\overline{s} P_{R(L)}  b\right) \left(\overline{\ell} \gamma_5 \ell\right)\,.
\end{align}
Matching to the singlet Lagrangian, the corresponding Wilson coefficients, $C_{i}$ with $i=S,P$, can be identified:
\begin{align}
\label{eq:LWET}
    \mathscr{L}_{a_{\pm}}^{\rm eff}  \supset \frac{(C_e^{\pm})_{\ell \ell} }{ m_{a_{\pm}}^2} \bigg[  
     (C_d^{\pm})_{bs} \mathcal{O}^\prime_{S(P)}\pm(C_d^{\pm})_{sb}  \mathcal{O}_{S(P)}\bigg]\,.
\end{align}

Instead, in the $m_{a_\pm} < m_B$ regime, the momentum dependence must be kept in the Wilson coefficients, which read as above upon the replacement ${1/m_{a_\pm}^2\to - 1/(q^2-m_{a_\pm}^2+\text{i}\Gamma_{a_\pm} m_{{a_\pm}})}$, $\Gamma_{{a_\pm}}$ denoting the singlet decay width. If furthermore ${a_\pm}$ lies within the energy window where the semileptonic decays, and consequently the LFU ratios, are determined, it can be produced \textit{on-shell} in $B_q\to X_s \ell^+ \ell^-$ decays. In this case, the narrow width approximation applies and the following expression holds:
\begin{align}
\label{eq:BRwSmearing}
    \mathcal{B}(B_q\to X_s \ell^+ & \ell^-)  =\mathcal{B}_\text{SM}(B_q\to X_s \ell^+\ell^-) \\
    & +\mathcal{B}(B_q\to X_s a_\pm)\mathcal{B}(a_\pm\to\ell^+\ell^-) \nonumber\,,
\end{align}
with
\begin{align}
     \mathcal{B} &(B\to  K a_\pm)\,\tau_B^{-1} =\, \\
    & \frac{ \left| (C_d^{\pm} )_{sb} \pm (C_d^{\pm})_{bs} \right|^2}{64  \pi M_B^3} \frac{(a_0^{B\to K})^2}{(m_b - m_s )^2}   \left( M_B^2 - M_K^2 \right)^2 \lambda^{1/2}_{BKa_\pm} \,;\nonumber
    \end{align}
    \begin{align}
      \mathcal{B}&(B\to K^* a_\pm)\,\tau_B^{-1}=\,  \\
      &  \frac{\left| (C_d^{\pm} )_{sb} \mp (C_d^{\pm})_{bs} \right|^2}{64  \pi M_B^3 } \frac{(a_0^{B\to K^*})^2 }{(m_b + m_s )^2} \lambda^{3/2}_{B K^* a_\pm} \,;\nonumber\\
      \label{eq:Bs}
     \mathcal{B} & (\bar{B}_s\to \phi a_\pm)\,\tau_{B_s}^{-1}=  \\ &  \frac{1}{1-y_s} \frac{\left| (C_d^{\pm} )_{sb} \mp (C_d^{\pm})_{bs} \right|^2}{16  \pi M_{B_s}^3 } \frac{(a_0^{B_s\to \phi})^2}{(m_b + m_s )^2}\lambda^{3/2}_{B_s K^* a_\pm} \,,\nonumber
\end{align}
where $\lambda_{B, K, a_\pm} \equiv \lambda(M_B^2, M_K^2, m_{ a_\pm}^2)$ is the K\"{a}ll\'en triangle function, $\tau_{B_q}$ are the $B$-meson lifetimes 
and $a_0^{B\to X_s}$~\cite{Bailey:2015dka,Bharucha:2015bzk} are the scalar form factors associated to the ${B_q\to X_s}$ transitions evaluated at ${q^2 = m_{a_\pm}^2}$. In Eq.~\eqref{eq:Bs}, ${y_s\equiv \Delta\Gamma_{B_s}/2\Gamma_{B_s}=0.0640 \pm 0.0035}$~\cite{ParticleDataGroup:2020ssz} takes into account the CP-oscillation of the $B_s$-meson. The previous relations imply, in particular, that
 \begin{align}
        \frac{\mathcal{B}(\bar{B}_s\to \phi a_\pm)}{\mathcal{B}(B\to K^* a_\pm)} &=\frac{1}{1-y_s}  \frac{\tau_{B_s}}{\tau_B}\left(\frac{M_B}{M_{B_s}}\right)^3 \left(\frac{a_0^{B_s\to\phi}}{a_0^{B\to K^*}}\right)^2 \nonumber\\
        & \approx (1.14 \pm 0.24) \,,
        \label{eq:BRphi}
    \end{align}
where the last line applies to $m_{a_\pm}^2 \in \left[1.1,6.0\right]$ GeV$^2$.

After production, the singlet can decay into fermion pairs. For the leptonic decay modes, we obtain:
\begin{equation}
    \frac{\Gamma (a_\pm \to \ell^+ \ell^-)}{m_{a_\pm}} =  \dfrac{(C^{\pm}_e )_{\ell\ell}^2}{8\,\pi} \left(1-\dfrac{4m_\ell^2}{m_{a_\pm}^2}\right)^{3/2 \,(1/2)}.
\end{equation}

%%%%%%%%%%%%%%%%%%%%%%%%%%
\section{Phenomenological analysis}~\label{sec:pheno}
In what follows, we explore the impact of an on-shell $a_\pm$ exchange to the neutral $B$-anomalies. As in this case interference effects are negligible, the electron channel plays the leading role in the analysis. We therefore set $\mathcal{B}(a_\pm\to e^+ e^-) = 1$. 

Our exploration of the singlet solutions to the anomalies is two-fold. First, we present predictions for several observables associated to the $B\to K^{(*)} e^+ e^-$ distributions assuming that $a_\pm$ explains  $R_{K^{(\ast)}}$ within $2\sigma$. The aim of this study is to quantify the experimental improvement required to probe the most interesting regions of the singlet parameter space.
Secondly, we translate that parameter space into the allowed range of singlet couplings and mass in order to infer potential constraints from other data. 

\subsection{Impact on $B \to K e^+ e^-$ observables}\label{sec:BtoK}
\renewcommand{\arraystretch}{1.6}
\begin{table*}[t!] \centering \resizebox{0.95\textwidth}{!}{
\begin{tabular}{|c|c|c|c|c|c|c|c|}
\hline 
$m_{a_\pm}$\,[GeV] & $\mathcal{B} (B\to K e^+ e^-)\,[10^{-7}]$ &$\mathcal{B} (B\to K^* e^+ e^-)\,[10^{-7}]$ & $F_L(B\to K^* e^+ e^-)[10^{-1}]$ & $A_{FB} (B\to K^* e^+ e^-)[10^{-2}]$ 
\\ \hline 
 1.5 & $[1.8 \pm 0.3,\,2.2\pm 0.4]$& $[2.6\pm0.4,4.5\pm 0.6]$   & $[4.0\pm0.2,6.9\pm0.4]$ & $[0.5\pm1.6,0.9\pm2.7]$ 
\\ \hline 
0.6 & SM-{\it like} &  $[1.4\pm0.2,2.4\pm 0.4]$ &  $[1.7\pm0.3,2.9\pm0.5]$ & $[-8.9\pm0.7,-5.1\pm0.4]$
\\ \hline
\end{tabular}}
\caption{\em On-shell NP predictions for $b\to s e^+ e^-$ observables involved in charged and neutral $B$-meson decays at $\mu = m_b$. Different $q^2$ cuts are considered: $\left[1.1,\,6.0 \right]$ GeV$^2$ and $\left[0.045,\, 1.1\right]$ GeV$^2$ in the first and second lines, respectively. The range of values span the compatibility with $R_{K^{(\ast)}}$ data within $2\sigma$.}
\label{tab:obs}
\end{table*}

If the NP couples only to electrons and reduces the $R_K$ tension within the SM, it must also produce an effect on $\mathcal{B}(B\to K e^+ e^-)$; see Tab.~\ref{tab:obs}. The majority of shifts produced on this observable, which are in this case independent of whether the NP is off- or on-shell, are within the SM uncertainty; see App.~\ref{app:SM}. An improvement of $\mathcal{O}(10\%)$ in the experimental bound would allow the probing of the present setup; see Eq.~\eqref{eq:BR-1}.

The angular observables associated to the $B\to K e^+ e^-$ distribution~\cite{Bobeth:2007dw},
\begin{equation}
    \frac{1}{\Gamma^\ell}\frac{\text{d}\Gamma^\ell}{\text{d} \cos{\theta} } =\frac{1}{2} F_H^\ell + \frac{3}{4}(1- F_H^\ell)\left(1-\cos^2{\theta}\right)  + A_{\rm FB}^\ell \cos{\theta}\,,
\end{equation}
provide another straightforward route to test the NP hypothesis being particularly sensitive to its nature. In the SM and similarly for new vector/axial interactions, $F_H^\ell \propto m_\ell^2$ such that it is completely negligible in the electron channel: $F_H^e < 10^{-6}$. In turn, $A_{FB}^\ell$ is zero at tree level in the SM, as it requires non-vanishing scalar, or pseudoscalar together with tensor, interactions. These arise in the SM only from higher dimensional operators or loop corrections inducing, even at NLO, a negligible $A_{FB}^\ell \sim m_\ell^2 / m_W^2$\cite{Bobeth:2007dw}.
Note that this couple of observables is normalized to the decay rate so that, as $R_K$, they are expected to be cleaner observables than $\Gamma^\ell$.

Being null tests of the SM, these angular observables could therefore provide a powerful probe of NP even if the contribution of the latter is small. In fact, ${F_H\propto |C_{{S,P}}|^2}$, hence being significantly enhanced if generated by a scalar (S) or pseudoscalar (P) exchange. Using the model-independent relation~\cite{Bobeth:2007dw}
\begin{equation}
R_K \left(1- F_H^\mu - C^{\rm SM} + \frac{F_H^e}{R_K}\right) = 1\,,
\label{eq:FHe}
\end{equation}
where $C^{\rm SM}$ remains unaltered by the presence of S/P interactions, we find that
\begin{equation}
\label{eq:FH}
    F_H^e = 1 - R_K (0.99922 \pm 0.00029)\,.
\end{equation}
Note that $F_H^\mu$ remains SM-like in the hypothesis underlying this work.
Using Eq.~\eqref{eq:FH}, we find that the solutions compatible with the data on $R_K$ in our setup produce values of $F_H^e \lesssim 0.3$, which are one order of magnitude larger than even $F_H^\mu$ in the SM~\cite{Bobeth:2007dw}; see Fig.~\ref{fig:corr}. No data currently available can probe these large SM deviations.

On the other hand, $A_{FB}^e < 0.1\%$ even for the largest values of the Wilson coefficients that can accommodate the $R_K$ anomaly. For this reason, the contributions from $\mathcal{O}_{S,P}$, generated by scalars of different CP-charge (see Eq.~\eqref{eq:LWET}), would remain undetermined by the proposed angular analysis.

\subsection{Impact on $B \to K^\ast e^+ e^-$ observables}
\begin{figure*}[htb!]
\centering
\includegraphics[width=0.4\textwidth]{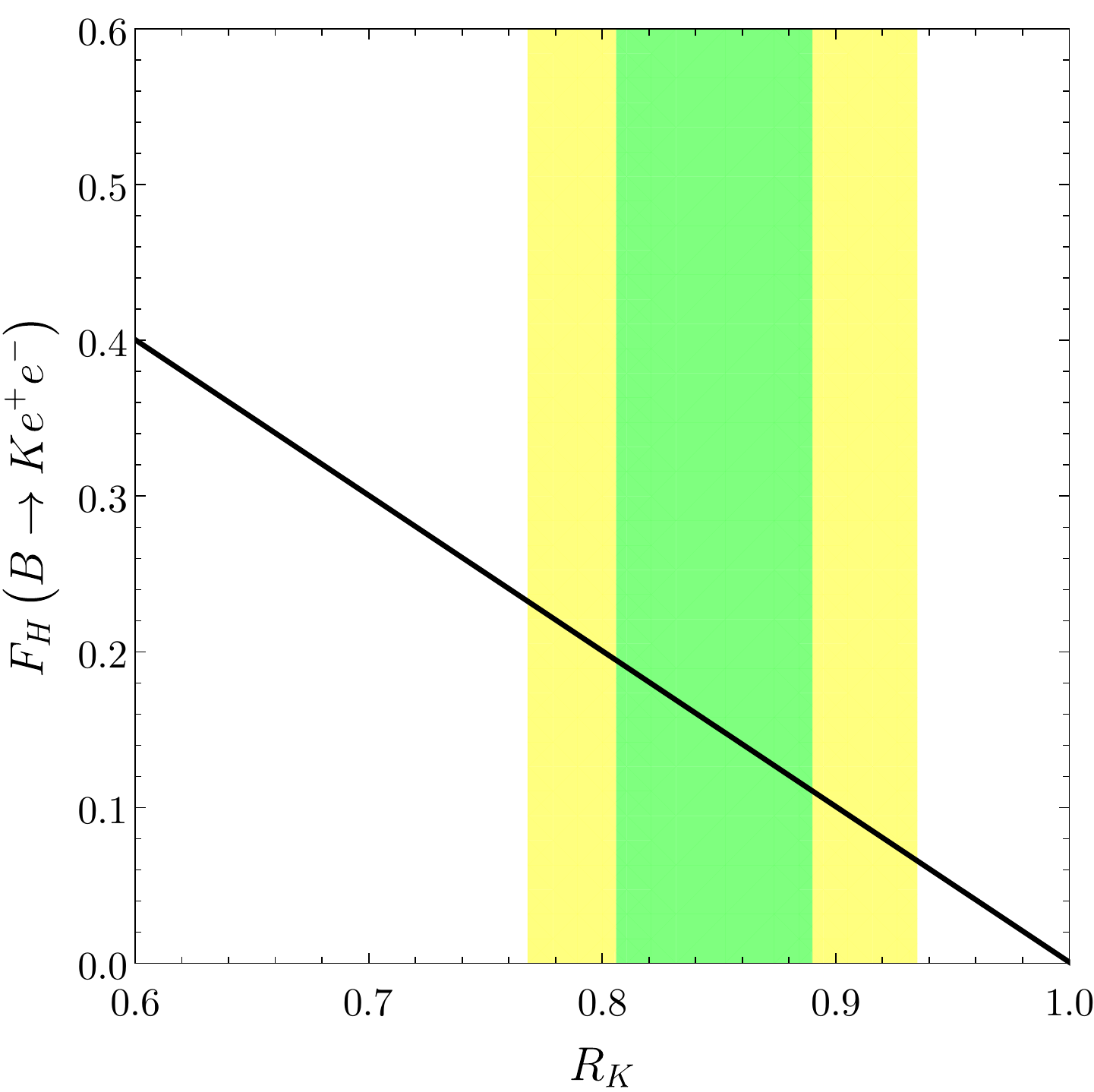}
\includegraphics[width=0.4\textwidth]{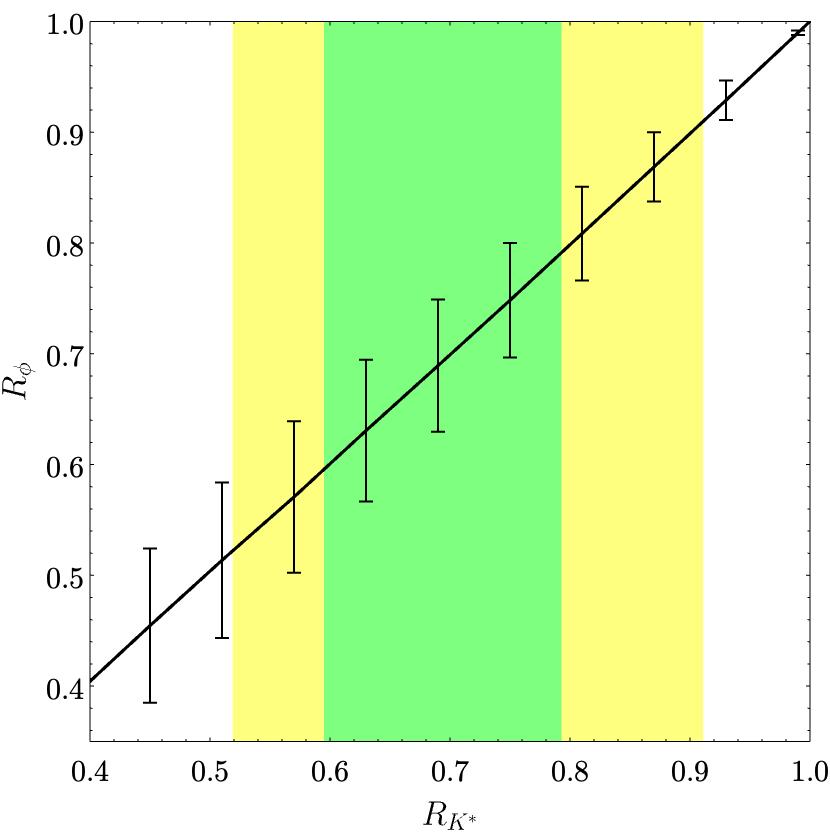}
\caption{\em Correlations between $b\to s e^+ e^-$ observables assuming the presence of (pseudo)scalar NP only in the electron sector: 
$R_K$ vs. $F_H^e$ (left) and $R_{K^\ast}$ vs. $R_\phi$ (right). The green (yellow) shaded area denotes the $1\sigma\,(2\sigma)$ experimental error range of $R_{K^{(\ast)}}.$}
\label{fig:corr}
\end{figure*}

The $B\to K^* e^+ e^-$ distribution is much richer than the one just analysed, being described by several angular coefficients~\cite{Altmannshofer:2008dz}:
\begin{equation}\label{eq:d4Gamma}
  \frac{\text{d}^4\Gamma^{ \ell*}}{\text{d}q^2\, \text{d}\cos\theta_l\, \text{d}\cos\theta_{K^*}\, \text{d}\phi} =
   \frac{9}{32\pi} I^\ell(q^2, \theta_l, \theta_{K^*}, \phi)\,,
\end{equation}
where
\begin{align} \label{eq:angulardist}
  I^\ell(q^2 & , \thl, \thK, \phi) = 
      I_1^s \sin^2\thK + I_1^c \cos^2\thK \\
      & + (I_2^s \sin^2\thK + I_2^c \cos^2\thK) \cos 2\thl
\nonumber \\       
    & + I_3 \sin^2\thK \sin^2\thl \cos 2\phi 
      + I_4 \sin 2\thK \sin 2\thl \cos\phi 
\nonumber \\       
    & + I_5 \sin 2\thK \sin\thl \cos\phi
\nonumber \\      
    & + (I_6^s \sin^2\thK +
      {I_6^c \cos^2\thK})  \cos\thl \nonumber \\
    &  + I_7 \sin 2\thK \sin\thl \sin\phi
\nonumber \\ 
    & + I_8 \sin 2\thK \sin 2\thl \sin\phi
      + I_9 \sin^2\thK \sin^2\thl \sin 2\phi\,. \nonumber
\end{align}
The coefficients $I_i$ (with the flavour superscript implicit) are functions of $q^2$ only. Particularly, the pseudoscalar operators affect $I_1^c$ while the scalar ones affect in addition $I_5$, $I_6^c$ and $I_7$ and therefore the observables defined upon these variables such as $\text{d}\Gamma^{ \ell*}/\text{d}q^2,\,P_5^{\ell\prime},\,P_6^{\ell\prime},\,F_L^\ell$ and $A_{FB}^{ \ell*}$~\cite{Altmannshofer:2008dz,Descotes-Genon:2013vna}. Among these, only the last two observables are normalized to $\Gamma^{\ell*}$. Even so, the $P^{\ell \prime}$ are considered to be largely free from hadronic uncertainties~\cite{Descotes-Genon:2013vna}. 

As in the previous case, the electrophilic NP that accommodates the $R_{K^*}$ data must also impact ${\mathcal{B} (B\to K^* e^+ e^-)}$; see Tab.~\ref{tab:obs}. Only the smallest values of this branching ratio compatible with the anomaly solutions are within the SM uncertainties at $1\sigma$. An improvement of $\mathcal{O}(30\%)$ in the measurement would be required to probe the full interval; see Eq.~\eqref{eq:BR-2}.

Furthermore, it is clear that any deviation in $R_{K^*}$ should propagate to other LFU ratios sensitive to the $b\to s e^+ e^-$ transition, such as $R_\phi$. Using Eq.~\eqref{eq:BRphi}, we find that if $a_\pm$ is responsible for the deviations found in $R_{K^*}$ at $1\sigma$, then $R_\phi \in [0.53,0.84]$ should be observed. The expected correlation between the two observables is shown in Fig.~\ref{fig:corr}.

Importantly, the \textit{on-shell} exchange of the particle can affect differently ${\mathcal{B}(B_s\to e^+ e^-)}$ and $\mathcal{B}(B\to K^* e^+ e^-)$ as long as the resonant peak is sufficiently separated from that of the $B$-mesons: in such case, the on-shell enhancement allows sufficiently small fermion couplings to produce a sizable effect in $R_{K^*}$ on resonance while leaving almost no imprint in $\mathcal{B}(B_s\to e^+ e^-)$ (we find that all corrections to the latter are within the SM uncertainty at $1\sigma$). This represents a striking difference with respect to the \textit{off-shell} regime where the NP corrections to the semileptonic rate required to explain the $R_{K^*}$ data are too large to comply with $B_s\to e^+ e^-$ constraints; see App.~\ref{sec:offshell}.

This \textit{on-shell} effect can only enhance observables sensitive to $C_{S,P}^2$; see App.~\ref{app:on-shell} for details. If an observable is instead sensitive only to the interference of the SM Wilson coefficients with $C_{S,P}$, the corresponding contributions are numerically irrelevant given the small values of the Wilson coefficients involved (antecipating the results in Tab.~\ref{tab:PS-Light}). Besides that, the S/P contributions to the angular observables via interference terms are suppressed by at least one power of $m_e/\sqrt{q^2}$, with $m_e$ denoting the electron mass. 
This implies that both $P_{5,6}^\prime$ remain SM-like. Particularly, $P_6^{e\prime}$ which is a null test of the SM is found to be $<10^{-8}$ including corrections from the NP.

Observables sensitive only to the (negligible) NP interference effects can however be corrected if normalized to the decay width $\Gamma^{\ell*}$, like $F_L^\ell$ and $A_{FB}^{\ell*}$, and  if measured in the resonant intervals  $[0.045,\,1.1]\cup[1.1,\,6.0]\,\text{GeV}^2$. More explicitly, a normalized observable $\mathcal{O}$ of this kind, in the presence of NP \textit{on-shell}, gets the following correction relatively to its SM values:
\begin{equation}
    \mathcal{O}=\mathcal{O}^\text{SM}\frac{1}{1+\left(\dfrac{\mathcal{B}^e_\text{NP}}{\mathcal{B}^e_{\text{SM}}}\right)}\,,
\end{equation}
where $\mathcal{B}^e_\text{NP}\equiv \mathcal{B}(B\to K^*a_\pm)$ in the underlying assumption of our study. Equivalently, as the same statement goes for $R_{K^*}$, we can write
\begin{equation}
\label{eq:Ocorr}
    \mathcal{O}=\mathcal{O}^\text{SM}\left(\dfrac{R_{K^*}}{R_{K^*}^\text{SM}}\right)\,.
\end{equation}
The corresponding predictions are given in Tab.~\ref{tab:obs}. Two mass benchmark points are studied\footnote{Note that the relation (couplings, mass) that accommodates the neutral anomalies is basically flat as long as $m_{a_\pm}$ lies within the bin window.}: $m_{a_\pm}=1.5$\,GeV (so that the singlet lies within the central $q^2$-bin) and $m_{a_\pm}=0.6$\,GeV (so that it lies within the low $q^2$-bin of $R_{K^\ast}$). As can be checked, both $F_L^e$ and $A_{ FB}^{e\ast }$ can only become smaller, up to $\mathcal{O}(50\%)$, than their SM values in order to comply with explanations to $R_{K^\ast}$ data.

Besides the previous ones, a particularly interesting observable proposed in Ref.~\cite{Matias:2012xw} to probe S/P NP is $M_2$, defined as
\begin{equation}
    \left<M_2\right>_{\rm bin}\equiv -\dfrac{\int_\text{bin} dq^2\, (I_1^c+I_2^c)}{\int_\text{bin} dq^2\, I_2^c}\,,
\end{equation}
where $I_1^c$ contains contributions $\propto|C_{S,P}|^2$.
If these coefficients are not generated by the UV, $M_2$ vanishes in the limit of massless leptons~\cite{Matias:2012xw}, which is 
in any case a good approximation to take here as $q^2 \gg m_e^2$ in the relevant energy bins. On the contrary, if such contributions are present, $M_2\neq 0$ even in the massless limit constituting a smoking-gun of S/P NP\footnote{Another observable dubbed $S_2 (I_7^2)$ was proposed in Ref.~\cite{Matias:2012xw} to probe new scalar interactions, which vanishes in their absence. Being $I_7 \sim {\rm Im}({\rm NP} \times {\rm SM})$, $S_2$ can be affected by the same on-shell enhancement as $M_2$. We do not consider it explicitly since it is a function of other eight angular coefficients that would have to be determined experimentally.}. Making use of the enhancement \textit{on-shell}, we find: 
\begin{align}
    &\left<M_2^e \right>_{0.045}^{1.1}\in[0.19\pm0.04,2.7\pm 0.6]\,,\\
    &\nonumber\left<M_2^e\right>_{1.1}^{6.0}\in[0.12\pm0.02 ,1.15\pm 0.23]\,,
\end{align}
for a singlet that lies within the low and central energy bins, respectively, and which explains the corresponding $R_{K^*}$ anomaly within $2\sigma$. In comparison, assuming the SM only, $\left<M_2^e\right>_{1.1}^{6.0}< $$10^{-6}$.

Finally, we remark that for non-scalar NP, the previous considerations do not hold. Since $e.g.$ a new spin-1 field can source the same operators as the SM, interference effects usually play the leading role and are sufficient to explain the $B$-anomalies~\cite{Alguero:2019ptt,Alok:2019ufo,Ciuchini:2019usw,DAmico:2017mtc,Aebischer:2019mlg}, unlike what is found in the scalar setup. In that case, Eq.~\eqref{eq:Ocorr} is no longer valid, neither the argument that the most sensitive angular observables would be those dependent on the NP$^2$.

\subsection{Additional new physics constraints}
\begin{table}[t!] 
\centering 
\resizebox{0.4\textwidth}{!}{
\begin{tabular}{|c||c|c|}
\hline 
& $a_+(1.5\,\text{GeV})$ & $a_+(0.6\,\text{GeV})$
\\ \hline
 \hline
$R_{K^{(*)}}$~bin & $[1.1,\,6]$&$[0.045,\,1.1]$\\ \hline
$\left|\left(C_d^+ \right)_{sb} + \left(C_d^+ \right)_{bs}\right|$ & $[1.0,2.1]\times 10^{-9}$ & 0 \\ \hline
$\left|\left(C_d^+ \right)_{sb} - \left(C_d^+ \right)_{bs}\right|$ & $[1.1,\,5.2] \times 10^{-9}$ & $[0.6,\,3.7] \times 10^{-9}$ \\ \hline
$\left|\left(C_e \right)_{ee}^+\right| $ & $2.0\times 10^{-5}$ & $2.9 \times 10^{-5}$\\ \hline
\end{tabular}}
\caption{\em Scalar singlet parameter space compatible with the $R_{K^{(*)}}$ anomalies within $2\sigma$ in the $q^2$ range indicated in the first line. The allowed range for the singlet-quark coupling is obtained by fixing the electron one by the prompt resonance condition. The values of the pseudoscalar couplings read the same upon the relabeling of ${|(C^+_d)_{sb} \pm (C^+_d)_{bs} | \to |(C^-_d)_{sb} \mp (C^-_d)_{bs} |}$ and ${|(C^+_e)_{\ell\ell}| \to |(C^-_e)_{\ell\ell}|}$.
}
\label{tab:PS-Light}
\end{table}

Across the range $2 m_\mu < m_{a_\pm} < \sqrt{6}$ GeV, the exotic singlet can provide an explanation not only to $R_K$ but also $R_{K^*}$. Interestingly, as shown in Tab.~\ref{tab:PS-Light}, this is accomplished for a single quark coupling being non-zero: $(C^\pm_d)_{bs} \sim 10^{-9}$. Note that in this case, assuming ${\mathcal{B}(a_\pm\to e^+e^-)\sim 100\%}$, the solutions are independent of the S/P nature of the NP; see Eq.~\eqref{eq:BRwSmearing}. We have explicitly checked, setting the singlet coupling to quarks to the maximum allowed value by measurements of $\mathcal{B}(B\to K^{(*)} e^+ e^-)$, in Eqs.~\eqref{eq:BR-2} and~\eqref{eq:BR-1}, that the $2\sigma\,(1\sigma)$ solutions to the anomalies require ${\mathcal{B}(a_\pm \to \mu^+ \mu^-) < 30\,(10)\%}$. This value becomes at least 50\% smaller taking into account the bounds set by the LHCb collaboration from dedicated searches of di-muon resonances~\cite{LHCb:2015nkv,LHCb:2016awg}. The assumed electrophilic character of the singlet is then a very robust assumption.

In the LHCb analyses of $R_{K^{(*)}}$, both the kaon and the dilepton system produced in $B\to K^{(*)}\ell^+\ell^-$ processes are required to come from the same vertex. This translates into a lower bound on the singlet coupling to electrons, $(C^\pm_e)_{ee}\gtrsim 10^{-5} $, to ensure it is prompt. The angular analyses presented in the previous section are independent of the specific value of the electron coupling. The latter can be however constrained by measurements of the anomalous magnetic moment of the electron $\Delta a_e$. In the limit $m_{a_\pm} \gg m_e$, the leading order (1-loop) contributions of $a_\pm$ to this observable read, respectively: 
\begin{equation}
\label{eq:Sg-2}
    \Delta a_e^{a_+} \approx (C_e^+)_{ee}^2 \dfrac{m_e^2}{8 \pi^2 m_{a_+}^2} \left( \log \left( \dfrac{m_{a_+}^2}{m_e^2} \right) - \dfrac{7}{6} \right) \,;
\end{equation}
\begin{equation}
\label{eq:Spg-2}
     \Delta a_e^{a_-} \approx - (C^-_e)_{ee}^2 \dfrac{m_e^2}{8 \pi^2 m_{a_-}^2} \left( \log \left( \dfrac{m_{a_-}^2}{m_e^2} \right) - \dfrac{11}{6}  \right) \,.
\end{equation}
These expressions agree with the results in Refs.~\cite{Giudice:2012ms,Queiroz:2014zfa}. In the relevant parameter space presented in Tab.~\ref{tab:PS-Light}, the NP contributions to $\Delta a_e$ are however $\lesssim 10^{-16}$, and hence 3 orders of magnitude smaller than the uncertainties of the experimental measurements based on Caesium atoms, $\Delta a_e^{\text{Cs}} =(-88\pm 36)\times 10^{-14}$~\cite{Parker:2018vye}, as well as those based on Rubidium atoms, $\Delta a_e^{\text{Rb}}  =(48\pm 30)\times 10^{-14}$~\cite{rubidium}.

\begin{figure}[t]
    \centering
    \includegraphics[scale=0.58]{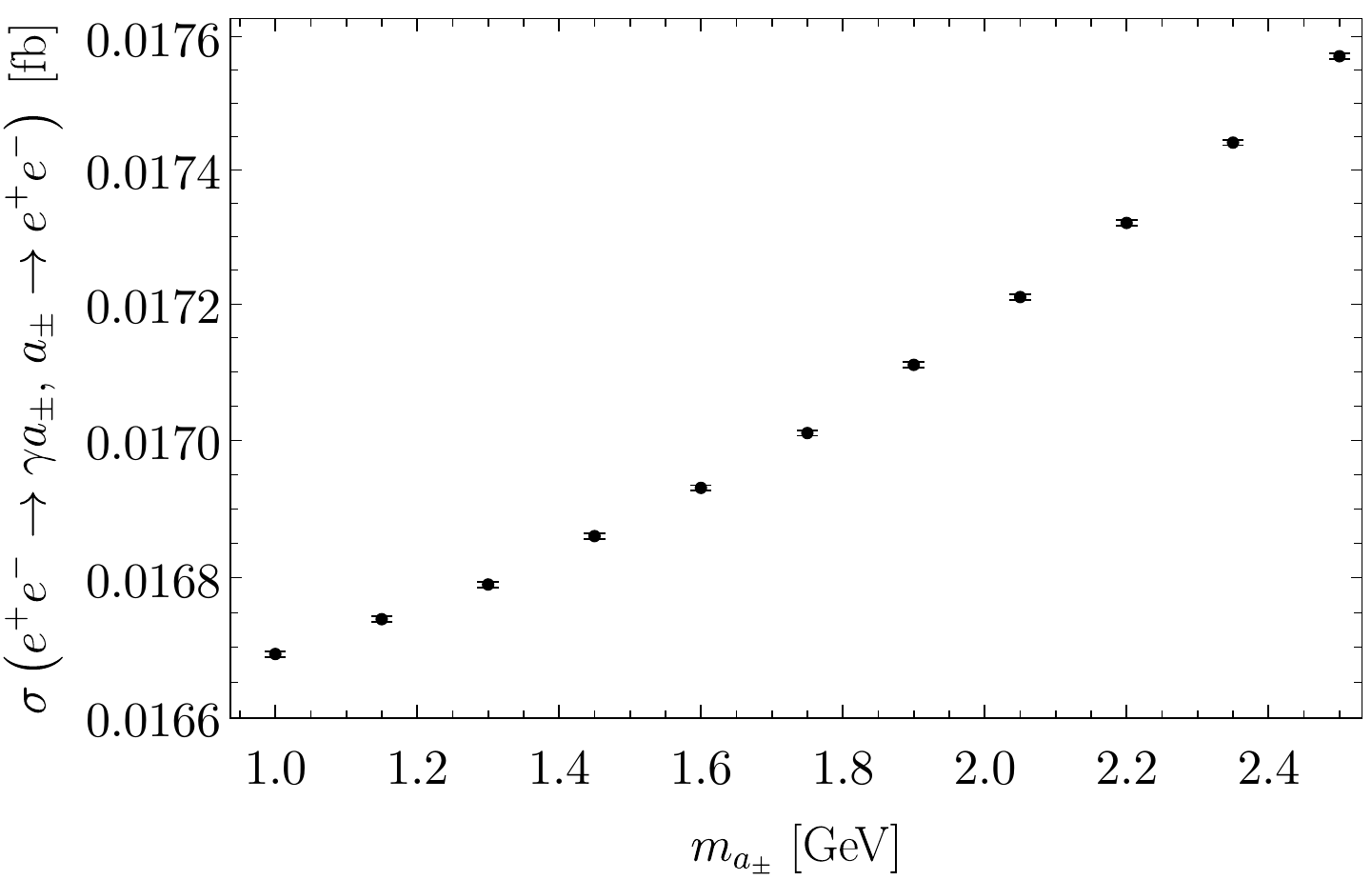}
    \caption{\em Production cross section of the electrophilic (pseudo)scalar at $e^+ e^-$ colliders with ${\sqrt{s}=10.58}$\,GeV, as a function of the mass.
    The electron coupling is set by the prompt condition; see Tab.~\ref{tab:PS-Light}.
    The results have been obtained with \textsc{MadGraph5\_\text{a}MC$@$NLO\,{\small v3}}~\cite{Alwall:2014hca,Frederix:2018nkq} with an UFO model obtained by means of  \textsc{FeynRules}~\cite{Alloul:2013bka}.
    Events were generated for a photon pseudo-rapidity $|\eta_\gamma| < 1.55$ corresponding to ${|\cos{\theta_\gamma}| < 0.91}$. (This approximately spans the acceptance region of the BaBar calorimeter~\cite{Essig:2009nc}.)
    }
    \label{fig:eeXS}
\end{figure}

On the other hand, searches in electron-positron colliders could potentially set important constraints on the electron coupling and therefore on the singlet lifetime, namely via the process ${e^+ e^- \to \gamma a_\pm,\,a_\pm\to e^+ e^-}$. The latter was analysed by BaBar in the relevant energy range in the context of dark photon models~\cite{BaBar:2014zli}. While a recast of the corresponding bounds is non-trivial as the interpretation takes into account both the electron and muon channels, it is reported that the $90\%$ CL limits on the cross section typically reach the level of $\mathcal{O} (1-10)$ fb. As can be seen in Fig~\ref{fig:eeXS}, for the minimum electron coupling compatible with the $R_{K^{(*)}}$ anomalies, the corresponding cross section is well beyond that reach. Belle II~\cite{Belle-II:2018jsg} could potentially set stronger constraints on the process of interest, given the large amount of data expected to be collected by the detector. Assuming conservatively the same experimental efficiency reported by BaBar, we can roughly estimate the observation of $\mathcal{O}(100)$ signal events at Belle II with an integrated luminosity of $50\,\text{ab}^{-1}$. 

Finally, as firstly pointed in Ref.~\cite{Bonilla:2022qgm}, there is the possibility that a \textit{golden} singlet close to the $1.1\,\text{GeV}^2$ threshold accommodates the anomalies in all bins explored so far. We dedicate the next section to studying this case. Otherwise, the compatibility of explanations to the central bin anomalies with those addressing the tension in the low-$q^2$ bin cannot be attained via an on-shell particle. The data on this low bin alone can be explained at $2\sigma$ by a singlet lying within the $[0.045,\,0.15] \cup [0.25,\,0.7]\,\text{GeV}^2$ bins of $q^2$; the corresponding allowed range of couplings for a 600 MeV resonance is presented in Tab.~\ref{tab:PS-Light}, for illustration purposes. Other mass values within $[0.045,\,1.1]\,\text{GeV}^2$ are excluded by binned measurements of ${\text{d}\mathcal{B}/\text{d} q^2 (B\to K^* e^+ e^-)}$; see  Eqs.~\eqref{eq:BR1}-\eqref{eq:BR-3}.

%%%%%%%%%%%%%%%%%%%%%%%%%%
\section{Golden scalar exchange}~\label{sec:golden}
\begin{figure*}[t!] 
\centering
\subfigure[\em Quark-coupling parameter space. \label{fig:fitQuarks}]
{\includegraphics[width=0.42\textwidth]{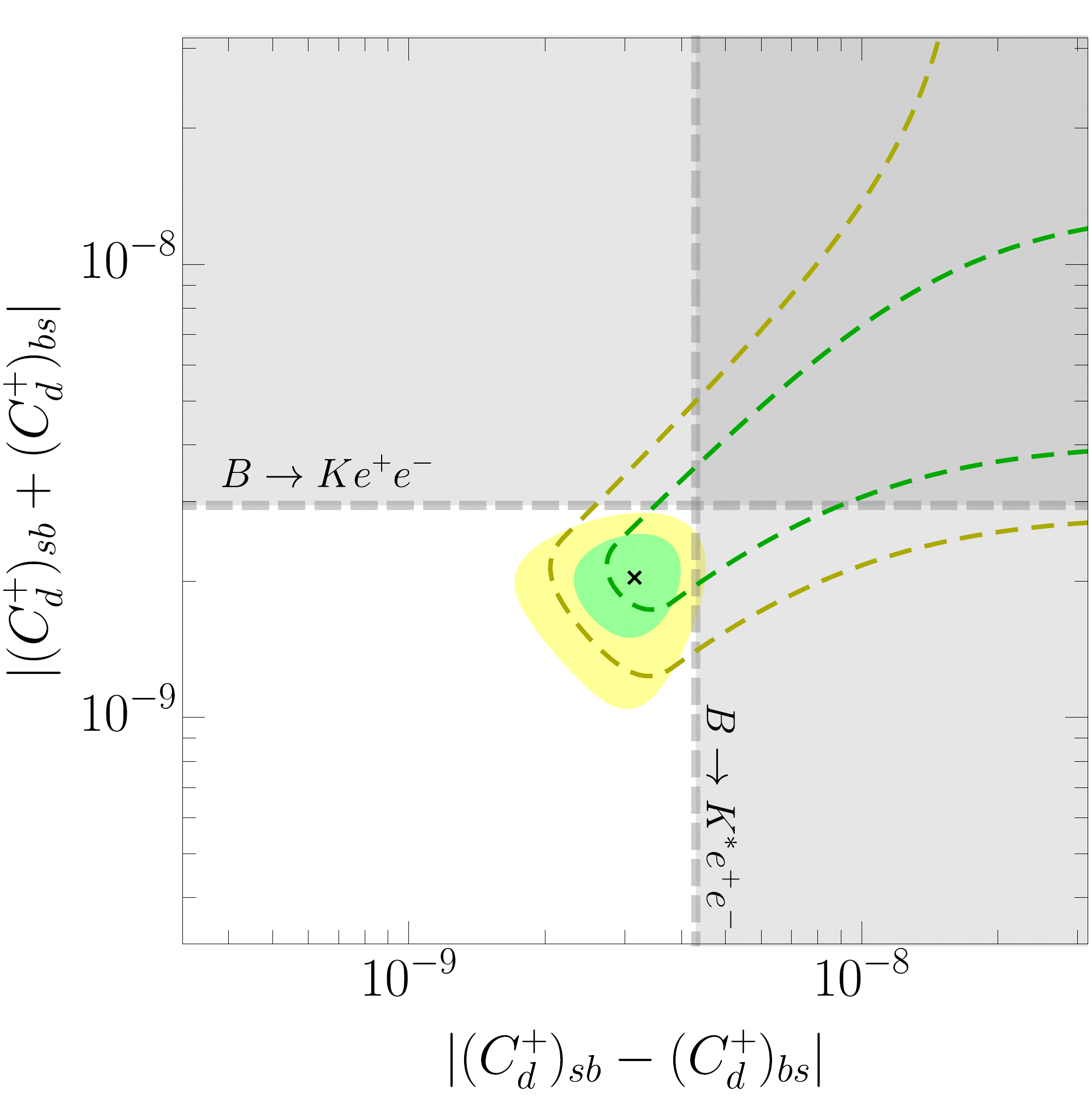}}\quad
\subfigure[\em Lepton-coupling parameter space. \label{fig:fitLeptons}]
{\includegraphics[width=0.42\textwidth]{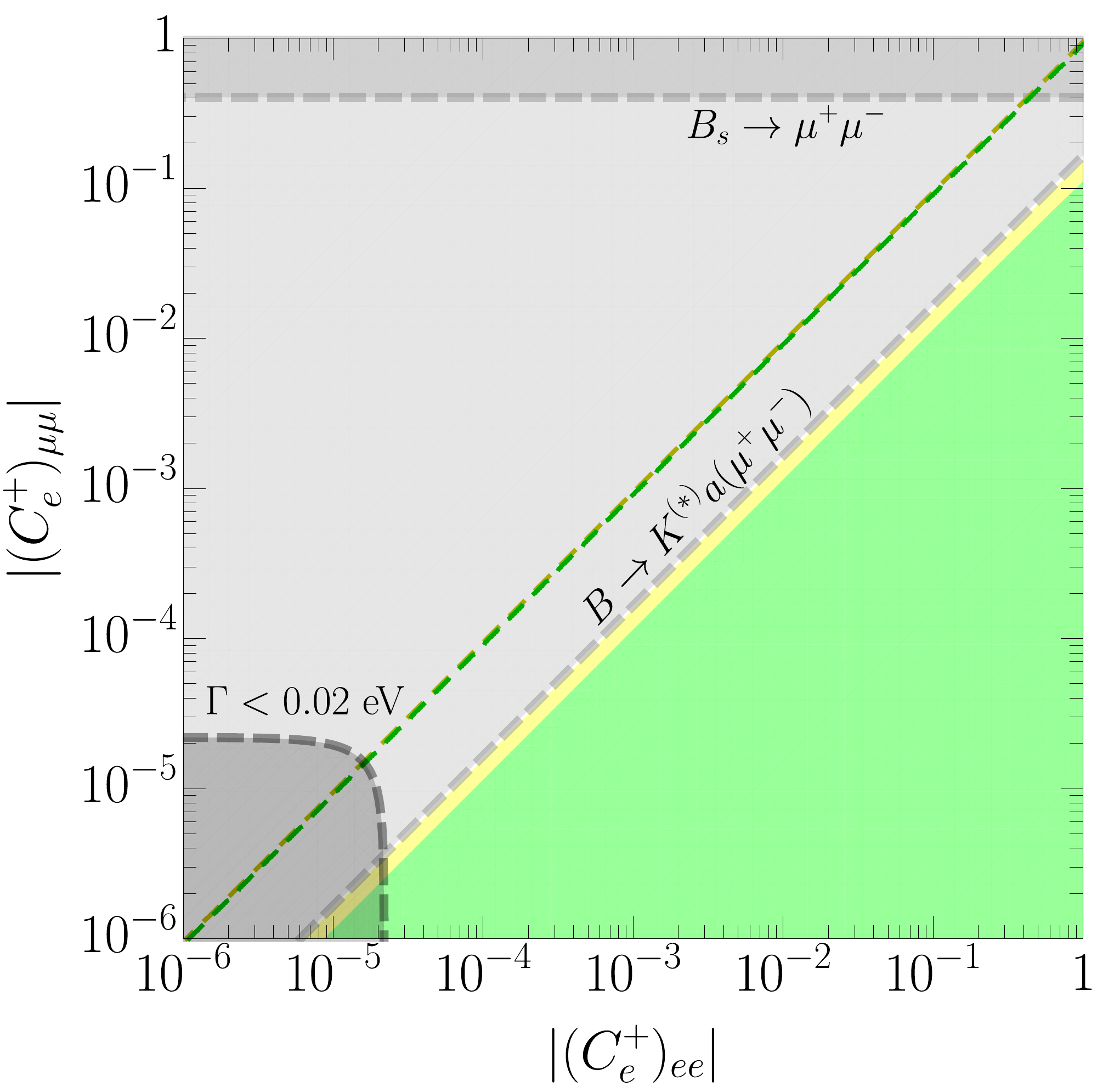}}
\caption{\em Preferred regions of the golden scalar parameter space where all $R_{K^{(*)}}$ anomalies can be solved simultaneously; see the text for details. Plots for the pseudoscalar are identical upon the relabeling of the axes: $|(C^+_d)_{sb} \pm (C^+_d)_{bs} | \to |(C^-_d)_{sb} \mp (C^-_d)_{bs} |$ and $|(C^+_e)_{\ell\ell}| \to |(C^-_e)_{\ell\ell}|$.
}
\label{fig:fit}
\end{figure*}

To the threshold solutions, experimental effects play an important role as, in consequence of the fact that the di-lepton mass resolution at LHCb is imperfect, resonant effects from a particle close to the bin thereshold can be partially smeared back into the measured $q^2$-bin.
We have included these effects by means of a gaussian function sensitive to the extremes of the relevant energy window~\cite{Altmannshofer:2017bsz}, in order to find the statistical significance of the three bin anomaly solution, for $m_{a_\pm} \approx \sqrt{1.1}$\,GeV. We quantified this solution including also a non-zero muon coupling to infer the impact of the latter on the phenomenological analysis.

With this aim, we have constructed a $\chi^2$-function based on a gaussian likelihood including the two most significant bins of $R_{K^*}$, $R_K$, the bounds on ${\mathcal{B} (B\to K^{(*)} e^+ e^-)}$ and ${\mathcal{B} (B_s\to e^+ e^-)}$; see section~\ref{sec:obs}. 
Regarding the semileptonic bounds, we use the measurements provided by the Belle collaboration (see App.~\ref{sec:obs}), to avoid experimental correlations between different analyses using the same dataset from which $R_{K^{(\ast)}}$ is determined. When including observables measured by the same experiment, we make a rough guess of the correlations by assuming the systematic uncertainties to be fully correlated. All experimental errors of observables entering into the fit are symmetrized. 
To remain under a conservative approach, we assume a $100\%$ correlation for the theoretical errors associated to the branching ratios quoted above (since, for example, the same CKM elements are used for their computation). 
We have neglected the NP contribution to the theory uncertainties, since their effects are small in comparison to the SM expectations. 
Overall, we have checked that the described procedure has a small impact on the resulting fit.

Together with the previous observables, the constraint on $F_L^e$ presented in Eq.~\eqref{eq:FLconstraint} is included in the fit, since this angular observable is measured in a bin where the $\sqrt{1.1}$ GeV resonance is localized (otherwise the effects are expected to be totally negligible, as argued in previous sections). Other data sensitive to the muon channel is included as well, namely measurements of $\mathcal{B} (B_s\to \mu^+ \mu^-)$~\cite{LHCb:2021vsc} and the bounds set on ${\mathcal{B} (B\to K^{(*)} a_\pm,a_\pm\to \mu^+ \mu^-)}$ by the LHCb collaboration~\cite{LHCb:2016awg,LHCb:2015nkv}. The results are presented in Figs.~\ref{fig:fitQuarks} and~\ref{fig:fitLeptons}, where the preferred regions of NP couplings at $68\%$ ($95\%$) CL are shown in green (yellow).

Fig.~\ref{fig:fitQuarks} spans the $|(C^+_d)_{sb} - (C^+_d)_{bs} | \text{ vs. } |(C^+_d)_{sb} + (C^+_d)_{bs} |$  parameter space, after profiling the $\chi^2$-function over the scalar-lepton couplings. The results of the fit including only the LFU ratios, that is, ignoring all other constraints, are also shown in the regions enclosed by dashed lines. The shaded grey regions are excluded at $95\%$ CL by the semileptonic constraints, under the assumption that the singlet decays $100\%$ into electrons. The black cross in this figure,  corresponding to $\Scale[0.95]{\left(|(C^+_d)_{sb} - (C^+_d)_{bs} |,\,|(C^+_d)_{sb} + (C^+_d)_{bs} |\right)\sim (3.2,\,2.1) \times 10^{-9}}$,
is the best-fit value in this case, with $\chi^2_{\rm min}/6 \approx 1.03$ ($10-4=6$ is the number of \textit{d.o.f.} in the fit). For the best-fit singlet couplings, we can also examine how much better the exotic singlet is at explaining the data than the SM. This is done by computing the pull, $\sqrt{\chi^2_{\rm SM} - \chi^2_{\rm  min}} \sim 5.02$.

In Fig.~\ref{fig:fitLeptons} we present instead the preferred regions of the $|(C^+_e)_{ee} | \text{ vs. } |(C^+_e)_{\mu\mu} |$ parameter space. The $\chi^2$-function is now profiled over the scalar-quark couplings; the preferred regions of the parameter space correspond to values of the latter close to the the best fit point in Fig.~\ref{fig:fitQuarks}. The dark gray region in this figure signals the regime where the resonance becomes non-prompt ($\Gamma_a < 0.02$ eV). While lepton couplings of the same order of magnitude (but satisfying ${|(C^+_{e})_{\mu\mu}| < |(C^+_{e})_{ee}|}$) are compatible with the fit including only the LFU ratios, we observe that ${ |(C^+_{e})_{\mu\mu}| \lesssim 10^{-1}|(C^+_{e})_{ee}|}$ in order to comply with the other constraints. 

In comparison, a (pseudo)scalar that couples with Higgs-like strength to the SM fermions, or an ALP comprising only shift-invariant interactions, has a reduced phase space in the plots above. Indeed, taking for example the ALP case, after re-expressing the derivative interactions with fermions in the Yukawa basis, the operators take the form $\mathbf{c} y^\psi  a \Psi_L \Phi \Psi_R /f_a$, with $\mathbf{c} < 1$ to make sense of a perturbative expansion. Due to the Yukawa suppression, the Wilson coefficients required in this framework to make the ALP decay promptly into electrons are $\mathbf{c}/f_a \gtrsim 10^{-2}\,\text{GeV}^{-1}$. Therefore, even for a cutoff relatively close to the EW scale, $\Lambda = 4\pi f_a \sim 1$\,TeV, the prompt condition implies that $\mathbf{c} \lesssim 1$, within but close to the perturbative bound.
Not having to beat the electron mass suppression, the generic singlet can naturally accommodate a larger phase space, with all couplings in Fig.~\ref{fig:fitLeptons} being perturbative even for larger values of $\Lambda$.

Ultimately, the analysis in this section shows that a 1 GeV spin-0 resonance can accommodate the anomalies observed in the three $q^2$-bins of $R_{K^{(\ast)}}$, at the level of $1\sigma$. This holds for masses within the range $[1.04,1.07]$\,GeV. Note that these values are stable against QED corrections~\cite{Bordone:2016gaq}, unlike the kinematic solutions close to the di-muon threshold analysed in Refs.~\cite{Altmannshofer:2017bsz,Bauer:2021mvw}. For smaller/larger mass values, such as those explored in the previous section, the smearing effect is irrelevant. 

It is interesting to note that the data on $B$-anomalies parameterized in the singlet model suggests a flavour structure that is very different for quarks and leptons: schematically, $C_e \sim \text{diag}(x,\,0,\,0)$ and $C_d \sim \text{diag}(0,\,0,\,y)$ in the UV could lead to the appearance of the couplings in Fig.~\ref{fig:fit}, with the non-diagonal $bs$ entry arising due to the running in the renormalization group~\cite{Chala:2020wvs}. 
Although the construction of a flavour model that produces these effects will be present in a work on progress,
power counting suggests that the loop and CKM suppressions inherited from this running allow $\mathcal{O}(x) \sim \mathcal{O}(y)$ as a solution. Additional non-diagonal entries in $C_d$ would also arise, however, suppressed by the previous factors, that could further constrain the singlet parameter space under a given flavour hypothesis.

%%%%%%%%%%%%%%%%%%%%%%%
\section{Conclusions}~\label{sec:conc}
The flavour physics arena has been under intense action since the so-called $B$-anomalies started to defy the SM-only scenario. In Ref.~\cite{Bonilla:2022qgm}, an interesting candidate to explain the data on neutral currents was identified: an axion-like particle that is produced on-shell in the semileptonic $B$-decays. In this work, we show that the exchange of a more general scalar singlet, comprising shift-breaking interactions with the SM fermions, offers a noticeably wider parameter space. On top of that, we study the pattern of experimental effects that the scalar solutions to the $R_{K^{(*)}}$ anomalies would leave in other observables, that goes well beyond the claim that scalar/pseudoscalar operators can be neglected in most $b\to s e^+ e^-$ data analyses (although the statement typically holds within the SMEFT~\cite{Altmannshofer:2008dz}).

Within this framework, we have argued that the data on the LFU ratios favors a singlet that decays sizably into electrons. In turn, (semi)leptonic and angular observables could distinguish the spin-$0$ nature of this candidate, as well as its \textit{on-shell} character. Summarizing our findings, we have shown that a (pseudo)scalar with a squared mass within the $\left[1.1,\,6.0 \right]\,\text{GeV}^2$ range can simultaneously explain the central $R_{K^{(*)}}$ bin anomalies within $2\sigma$, leading to: 
\begin{enumerate}
    \item Up to $\mathcal{O}(10-30)\%$ corrections in the exclusive $\mathcal{B}(B\to K^{(*)} e^+ e^-)$ with respect to the SM predictions;
    \item Negligible effects in the leptonic ${B_s \to e^+ e^-}$ decay;
    \item Enhancement of the observables ${F_H (B\to K e^+ e^-)}$ and $M_2 (B\to K^* e^+ e^-)$, which are in essence null tests of the SM, up to levels of $\mathcal{O}(0.1)$;
    \item Subleading contributions to all $P_i^{\prime} (B\to K^* e^+ e^-)$ observables, that are therefore expected to remain SM-like;
    \item Deviations in the $R_\phi$ ratio with respect to the SM prediction, reaching values $\gtrsim 0.5$;
    \item Corrections to all normalized observables associated to the $B\to K^* e^+ e^-$ distribution and which are only sensitive to the NP at linear order, such as $F_H$ and $A_{FB}$. These corrections are predicted to be proportional to those induced in $R_{K^{*}}$. 
\end{enumerate}
All these effects are expected to occur in the same bins where the $R_{K^{(*)}}$ anomalies are observed. While the previous observables could potentially also be able to differentiate between scalar/pseudoscalar interactions, we have found that, for the smallness of the fermion couplings involved, the CP nature of the new particle would remain undetermined, assuming that CP is a symmetry of the new physics sector.

Due to the on-shell enhancement, the electron coupling does not directly contribute to the effects stated above, but it is of major importance to guarantee that the resonance is prompt and that it decays mostly into electrons, a requirement to explain the $R_{K^{(*)}}$ anomalies. We have found that the impact of the new physics on $(g-2)_e$ data is negligible. On the other hand, improved sensitivity to the process $e^+ e^- \to a_\pm \gamma,\, a_\pm\to e^+ e^-$ at future electron-positron colliders could probe this prompt condition. The latter implies a cross section of the given process of $\mathcal{O}(10^{-2})$\,fb at $\sqrt{s}=10.58$\,GeV.

Not only combined solutions to the central anomalies are possible in this framework, but also a simultaneous explanation of the data measured in the low $q^2$-bin of $R_{K^*}$, at the level of $1\sigma$. This requires, however, a resonance very close to the ${1.1}\,\text{GeV}^2$  threshold. Solutions of this type could be straightforwardly tested in updated $R_{K^{(*)}}$ analyses with a different binning that avoids the overlap between the energy intervals, as suggested in Ref.~\cite{Bonilla:2022qgm}. On the other hand, the solutions to the central bin anomalies described above are expected to hold even for different choices of the bin extremes. Given the non-trivial pattern of effects left on other observables, these solutions could be probed with new data associated to the semileptonic distributions in the electron channel.

%%%%%%%%%%%%%%%%%%%%%%%%
\section*{Acknowledgements}
The authors are specially thankful to Belen Gavela and Luca Merlo for enlightening discussions and valuable feedback on the manuscript, as well as to Ilaria Brivio, Guilherme Guedes, Olcyr Sumensari and Jean-Loup Tastet for suggestions that enriched the scope of the work. MR thanks also MITP for the hospitality during the completion of this work and in particular Carla Benito, Javier Fuentes-Martín and Claudio Manzari for the interesting discussions. This project has received partial funding/support by the Spanish Research Agency (Agencia Estatal de Investigaci\'on) through the grant IFT Centro de Excelencia Severo Ochoa No CEX2020-001007-S and by the grant PID2019-108892RB-I00 funded by MCIN/AEI/ 10.13039/501100011033, by the European Union's Horizon 2020 research and innovation programme under the Marie Sk\l odowska-Curie grant agreement No 860881-HIDDeN. The work of J.B. was supported by the Spanish MICIU through the National Program FPU (grant number FPU18/03047).

\appendix
\renewcommand{\theequation}{\thesection.\arabic{equation}}

\section{Observables}\label{sec:obs}
The $b\to s e^+ e^-$ transition induced by a (pseudo)scalar particle exchange can impact several observables. Among those, the ones that have been determined experimentally and set the most important constraints in our setup are:

\begin{itemize}

\item The $R_{K}$ ratio measured in the central $q^2$-bin and $R_{K^*}$ in both low and central bins by the LHCb collaboration:
\begin{align}
&\left<{R_K}\right>_{1.1}^{6.0}=0.846^{+0.042}_{-0.039}{}^{+0.013}_{-0.012}~\text{\cite{LHCb:2021trn}}\,,\\
& \left<R_{K^*}\right>_{1.1}^{6.0} =0.69^{+0.11}_{-0.07}{}\pm0.05~~\text{\cite{LHCb:2017avl}}\,, \\
& \left<R_{K^*}\right>_{0.045}^{1.1} = 0.66^{+0.11}_{-0.07}\pm0.03
~\text{\cite{LHCb:2017avl}}\,,
\end{align}
where here and in the following only the most constraining intervals of $q^2\,[\text{GeV}^2]$ are shown.

\item The branching ratio $\mathcal{B} (B_s \to e^+ e^-)$ constrained to be $< 11.2 \times 10^{-9}$ at 95\% CL by the LHCb collaboration~\cite{LHCb:2020pcv};

\item The differential semileptonic branching ratio ${\text{d} \mathcal{B}^{e(\ast)}/\text{d} q^2\equiv \text{d}\mathcal{B}/\text{d} q^2( B \to K^{(*)} e^+ e^-)}$ determined by the LHCb collaboration in various energy bins. Integrating over the energy:
\begin{align}
\label{eq:BR1}
\left<\mathcal{B}^{e*}\right>_{0.0004}^{0.0500} &= (2.1 \pm 0.4) \times 10^{-7} ~\text{\cite{Bonilla:2022qgm}}
\,;\\
 \left<\mathcal{B}^{e*}\right>_{0.05}^{0.15} &= (2.6 \pm 1.2) \times 10^{-8} ~\text{\cite{Bonilla:2022qgm}} \,;\\
 \left<\mathcal{B}^{e*}\right>_{0.15}^{0.25} &= (6.1 \pm 6.2) \times 10^{-9} ~\text{\cite{Bonilla:2022qgm}} \,;\\
\left<\mathcal{B}^{e*}\right>_{0.25}^{0.40} &= (2.1 \pm 0.9 )\times 10^{-8}~\text{\cite{Bonilla:2022qgm}} \,;\\
  \left<\mathcal{B}^{e*}\right>_{0.4}^{0.7} &= (2.2 \pm 1.0)\times 10^{-8}~\text{\cite{Bonilla:2022qgm}}
\,;\\
\label{eq:BR-3}
 \left<\mathcal{B}^{e*}\right>_{0.7}^{1.0} &= (1.1 \pm 0.7)\times 10^{-8}~\text{\cite{Bonilla:2022qgm}}
\,;\\
 \label{eq:BR-1}
 \left<\mathcal{B}^e\right>_{1.1}^{6.0} & =(14.01^{+0.98}
_{ - 0.83} \pm 0.69) \times 10^{-8}~\text{\cite{LHCb:2019hip}} \,.
\end{align}

\item The Belle collaboration has provided complementary information on these decays, namely by providing measurements in the following bins:
\begin{align}
   \label{eq:BR-2}
 \left<\mathcal{B}^{e*}\right>_{1.1}^{6.0} &= (1.8\pm 0.6)\times 10^{-7} ~\text{\cite{Belle:2019oag}} 
\,; \\
\label{eq:BR-3}
 \left<\mathcal{B}^e\right>_{1.0}^{6.0} &=(1.66^{+0.32}
_{ - 0.29} \pm 0.04) \times 10^{-7}~\text{\cite{BELLE:2019xld}}\,.
\end{align}

\item The angular observable $F_L^e$, related to the $K^*$ polarization in the $B\to K^* e^+ e^-$ distribution, which was measured by the LHCb collaboration~\cite{LHCb:2015ycz}:
\begin{align}
\label{eq:FLconstraint}
\left<F^e_{L}\right>_{0.002}^{1.12} & = 0.16 \pm 0.06 \pm 0.03 \,.
\end{align}
For all angular observables, we do not consider previous measurements that average over electrons and muons since they cannot be applied in a straightforward way and are typically subject to larger uncertainties~\cite{BaBar:2006tnv}. 

\item The $P^{e\prime}_{i}$ observables associated to the ${B\to K^* e^+ e^-}$ decay, some of which have been measured by the Belle collaboration~\cite{Belle:2016fev}, namely
\begin{align}
\left<P^{e\prime}_{5}\right>_{1.0}^{6.0}& = - 0.22^{+0.39}_{-0.41} \pm 0.03 \,.
\end{align}
\end{itemize}

%%%%%%%%%%%%%%%%
\section{SM Inputs}
\label{app:SM}
All the SM inputs used in this work are listed below. The corresponding values are obtained with \texttt{Flavio}\cite{Straub:2018kue}. 
\textbf{\small 1.~Branching and flavour ratios}
\begin{align}
     \left<\mathcal{B}_\text{SM}(B_s^0\to \phi e^+e^-)\right>_{1.1}^{6.0} &= (2.64\pm 0.32)\cross 10^{-7}\,\\
      \left<\mathcal{B}_\text{SM}(B_s^0\to \phi \mu^+ \mu^-)\right>_{1.1}^{6.0} &= (2.63\pm 0.32)\cross 10^{-7}\,
      \end{align}
      \begin{align}
      \left<\mathcal{B}_\text{SM}^{e}\right>_{1.0}^{6.0} &= (1.75\pm 0.32)\cross 10^{-7}\,
      \end{align}
      \begin{align}
    \left<\mathcal{B}_\text{SM}^{e}\right>_{1.1}^{6.0} &= (1.71\pm 0.29)\cross 10^{-7}\,  \\
    \left<\mathcal{B}_\text{SM}^\mu\right>_{1.1}^{6.0} &= (1.71\pm 0.29)\cross 10^{-7}\,
    \end{align}
      \begin{align}
         \left<\mathcal{B}_\text{SM}^{e\ast}\right>_{1.1}^{6.0} &= (2.34\pm 0.34)\cross 10^{-7}\,\\
      \left<\mathcal{B}_\text{SM}^{\mu\ast}\right>_{1.1}^{6.0} &= (2.33\pm 0.36 )\cross 10^{-7}\,
      \end{align}
      \begin{align}
    \left<\mathcal{B}_\text{SM}^ {e\ast}\right>_{0.045}^{1.1} &= (1.29\pm 0.20)\cross 10^{-7}\, \\
     \left<\mathcal{B}_\text{SM}^{\mu\ast}\right>_{0.045}^{1.1} &= (1.22\pm 0.20)\cross 10^{-7}\, 
     \end{align}
      \begin{align}
      \left<R^\text{SM}_\phi\right>_{1.1}^{6.0} & =0.99644\pm0.00057 \\
      \left<R^\text{SM}_K\right>_{1.1}^{6.0} &=1.00078\pm 0.00029\,\\
     \left<R^\text{SM}_{K^*}\right>_{1.1}^{6.0} &=0.99644\pm0.00057 \,\end{align}
      \begin{align}
     \left<R^\text{SM}_{K^*}\right>_{0.045}^{1.1} & =0.926\pm0.004
\end{align}

\textbf{\small 2.~Angular observables}
\begin{align}
    \av{F_L^\text{SM}(B^0\to K^* e^+e^-)}_{1.1}^{6.0}&= 0.76 \pm 0.04\,\\
    \av{A_{FB}^\text{SM}(B^0\to K^* e^+e^-)}_{1.1}^{6.0}&= 0.01\pm 0.03\,
    \end{align}
    \begin{align}
        \label{eq:FHmu}
        \av{F_H^\text{SM}(B\to K \mu^+\mu^-)}_{1.1}^{6.0}&= 0.0239 \pm 0.0003\,\\[8pt]
    \av{F_H^\text{SM}(B^0\to K^* e^+e^-)}_{0.045}^{1.1}&= 0.31\pm0.05\,\\
    \av{A_{FB}^\text{SM}(B^0\to K^* e^+e^-)}_{0.045}^{1.1}&=-0.094\pm 0.007 \,
       \end{align}
    \begin{align}
\av{F_L^\text{SM}(B^0\to K^* e^+e^-)}_{0.002}^{1.120}&=0.18\pm0.04 \,
\end{align}

%%%%%%%%%%%%%%%
\section{On-shell enhancement} \label{app:on-shell}
The on-shell enhancement of a given observable can be understood from the narrow-width approximation, that is, the fact in the limit $\Gamma_{a_\pm}/m_{a_\pm} \ll 1$, the contribution from a tree level particle exchange can be approximated by
\begin{align}
\label{eq:enhancement}
    \left|\frac{\left(C_e\right)_{\ell \ell}}{(q^2-m^2)+i m\Gamma}\right|^2& =\frac{\left(C_e\right)_{\ell \ell}^2 }{m\Gamma}\frac{m\Gamma}{(q^2-m^2)^2+(m\Gamma)^2}\nonumber \\
    &\approx \frac{\pi \left(C_e\right)_{\ell \ell}^2}{m\Gamma}\delta(q^2-m^2)\,.
    \end{align}
where we have made use of the following identity:
\begin{equation}
\label{eq:delta}
    \delta(x)=\frac{1}{\pi}\lim\limits_{\epsilon\to 0}\frac{\epsilon}{x^2+\epsilon^2}\,.
\end{equation}
The $\Gamma$ dependence on the denominator makes the observable insensitive to the coupling $\left(C_e\right)_{\ell \ell}$ on resonance. The $a_\pm$ subscript is implicit.

The same type of enhancement can be obtained if an observable is sensitive to the imaginary part of the propagator squared:
\begin{align}
        \Im{\frac{\left(C_e\right)_{\ell \ell}}{(q^2-m^2)+i m\Gamma}}^2 &=\frac{\left(C_e\right)_{\ell \ell}^2}{m\Gamma}\frac{(m\Gamma)^3}{\left[(q^2-m^2)^2+(m\Gamma )^2\right]^2}\nonumber \\
        &\approx\frac{\pi\left(C_e\right)_{\ell \ell}^2}{2m\Gamma}\delta(q^2-m^2)\,. 
    \end{align}

On the contrary, no on-shell enhancement occurs in interference terms, as
\begin{align}
        &\frac{\left(C_e\right)_{\ell \ell}}{(q^2-m^2)+i m\Gamma} =\left(C_e\right)_{\ell \ell} \frac{(q^2-m^2)-i m\Gamma}{(q^2-m^2)^2+(m\Gamma )^2} \\
        &\approx \left(C_e\right)_{\ell \ell} \frac{(q^2-m^2)}{(q^2-m^2)^2+(m\Gamma )^2} -i\pi \left(C_e\right)_{\ell \ell}\delta(q^2-m^2)\,.\nonumber
    \end{align}

%%%%%%%%%%%%%%%
\section{Comparison with off-shell exchange}~\label{sec:offshell}

A combined explanation to the central bin anomalies via an off-shell (pseudo)scalar exchange is ruled out given the incompatibility of the $R_{K^*}$ solutions with $B_s \to e^+ e^-$ data~\cite{Bonilla:2022qgm}. Furthermore, the constraints imposed by measurements of the latter give very weak prospects to constrain the (pseudo)scalar parameter space via the $B \to K^* e^+ e^-$ angular observables~\cite{Altmannshofer:2008dz}. On the other hand, the effects in $B\to K e^+ e^-$ observables are expected to be similar to those reported in section~\ref{sec:BtoK}. Given that the effective operators contributing to each anomaly have opposite parity in our framework and that the $R_{K^{(\ast)}}$ data have different statistical significance, we identify in this section the (less interesting) parameter space where a heavy singlet can explain solely the $R_K$ anomaly. (For the complementary case of a very light resonance, no room is left for even one-anomaly solutions~\cite{Bonilla:2022qgm}.) 

\begin{table}[t!] 
\centering 
\resizebox{0.435\textwidth}{!}{
\begin{tabular}{|c||c|c|}
\hline 
& $a_+~(10\,\text{GeV})$ & $a_-~(10\,\text{GeV})$ 
\\ \hline
 \hline
$\left|\left(C_d^+ \right)_{sb} + \left(C_d^+ \right)_{bs}\right|$ & $3.3\times 10^{-5}$& --- \\ \hline
$\left|\left(C_e^+ \right)_{ee}\right| $ & $[5.8,\,12]\times 10^{-3}$ & ---\\ \hline
$\left|\left(C^-_d \right)_{sb} - \left(C^-_d \right)_{bs}\right|$ & --- & $1.1\times 10^{-4}$ \\ \hline
$\left|\left(C^-_e \right)_{ee}\right| $ & --- & $[1.7,\,3.7]\times 10^{-3}$\\ \hline
\end{tabular}}
\caption{\em Heavy singlet parameter space compatible with the $R_{K}$ anomaly within $2\sigma$. The allowed range for the singlet-electron coupling is obtained by fixing the quark coupling to its maximum allowed value by $\Delta M_s$ constraints.}
\label{tab:PS-Heavy}
\end{table}

In this high-mass regime, bounds from meson mixing provide the strongest constraints on the singlet coupling to quarks. Fixing the latter to the maximum value allowed by data at $2\sigma$~\cite{LHCb:2021moh} and using the values for the SM prediction obtained in Refs.~\cite{FermilabLattice:2016ipl,DiLuzio:2019jyq}, we identify the minimum values of the electron coupling that solve the $R_{K}$ anomaly within $2\sigma$; see Tab.~\ref{tab:PS-Heavy}. Solutions with larger significance are found for central values in the quoted interval; for instance, $|(C_e^\pm)_{ee}| \in [7.7\,(2.3),\,11\,(3.3)]\times 10^{-3}$ explain the $R_K$ data at the $1\sigma$ level.

The predictions for ${\Gamma (B\to K e^+ e^-)}$ presented in Tab.~\ref{tab:obs} hold for a heavy scalar that solves the $R_K$ tension coupling only to electrons, as well as the correlation $R_K$ vs.~$F_H^e$ shown in the left panel of Fig.~\ref{fig:corr}. A remark is however in order: to explain the $R_K$ data via an on-shell exchange, very small values of the fermion couplings combined are allowed (see Tab.~\ref{tab:PS-Light}), about six orders of magnitude smaller than those required for a particle that is exchanged off-shell. This can have an impact on other data such as $\Delta a_e$ that is analysed below. The fact that both cases generate the same effect in $F_H^e$ can be explained by the on-shell enhancement that makes $F_H^e$ much larger on resonance than off resonance, by a factor of $1/(C^\pm_e)_{ee}^2$; see App.~\ref{app:on-shell}.

In turn, the values of the electron coupling compatible with the $R_K$ solutions can be constrained by measurements of $\Delta a_e$. Using Eqs.~\eqref{eq:Sg-2} and~\eqref{eq:Spg-2}, together with the values presented in Tab.~\ref{tab:PS-Heavy}, we find:
\begin{align}
    &\Delta a_e^{a_+} \in  [2.1,\,8.9] \times 10^{-14}\,,\\
    &\nonumber\Delta a_e^{a_-} \in -[0.17,\,0.81] \times 10^{-14}\,.
\end{align}
Both these values are incompatible with the experimental measurement based on Caesium atoms within $2\sigma$ (but so is the SM)~\cite{Aoyama:2017uqe}.
Regarding the measurements with Rubidium atoms, both values are within the experimental uncertainty. Due to the discrepancy between the two measurements, plus the fact that the tree level singlet-photon coupling can arise in the singlet EFT (impacting $\Delta a_e$), such constraints should be taken with care.

%%%%%%%%%%%%%%%%%%%%%%%

\bibliographystyle{BiblioStyle}
\bibliography{bibliography}{}

\end{document}